\documentclass[apj]{emulateapj}
\slugcomment{Draft version 01/07/2008}
\usepackage{graphicx,amssymb,amsmath,times}

\def\swift{{\it Swift}/BAT}
\def\asm{{\it RXTE}/ASM}
\def\ariel{{\it Ariel\,V}/ASM}
\def\vela{{\it Vela\,5B}/ASM}
\def\numin{$\nu_1$}
\def\numax{$\nu_{m/2}$}
\def\nupeak{$\nu_\textrm{peak}$}
\def\cyg{Cygnus~X-1}

\def\mod{$A$}
\def\batse{{\it CGRO}/BATSE}

\def\postrial{${\cal P}_\textrm{post}$}

\begin{document}
\title{The variable super-orbital modulation of Cygnus X-1}

\shorttitle{The variable super-orbital modulation of Cygnus X-1}
\shortauthors{Rico}

\author{
Javier Rico\altaffilmark{1,2}
}

\altaffiltext{1} {Instituci\'o Catalana de Recerca i Estudis
Avan\c{c}ats (ICREA), E-08010 Barcelona, Spain. E-mail: jrico@icrea.cat}
\altaffiltext{2} {Institut de F\'{\i}sica d'Altes Energies (IFAE), Edifici Cn., Campus UAB, E-08193
Bellaterra, Spain. E-mail: jrico@ifae.es}

\begin{abstract}  
We study the super-orbital modulation present in the \cyg\ X-ray data,
usually attributed to the precession of the accretion disk and
relativistic jets. We find a new, strong, 326$\pm$2~d period
modulation starting in 2005, in \swift\ and \asm\ light curves
(LCs). We also investigate \vela\ and \ariel\ archival data and
confirm the previously reported $\sim$290~d periodic modulation, and
therefore confirming that the super-orbital period is not
constant. Finally, we study \asm\ LC before 2005 and find that the
previously reported $\sim$150~d period is most likely an artifact due
to the use of a Fourier-power based analysis under the assumption that
the modulation has a constant period along the whole data
sample. Instead, we find strong indications of several discrete
changes of the precession period, happening in coincidence with soft
and failed state-transition episodes. We also find a hint of
correlation between the period and the amplitude of the
modulation. The detection of gamma-rays above 100 GeV with MAGIC in
September 2006 happened in coincidence with a maximum of the
super-orbital modulation. The next maximum will happen between 2 and
14 of July 2008, when the observational conditions of
\cyg\ with ground-based Cherenkov telescopes, such as MAGIC and
VERITAS, are optimal.
\end{abstract}


\section{Introduction} 

Cygnus X-1\citep{Bowyer1965} is the best established candidate for a
stellar mass black-hole (BH). It is composed of a $21\pm
8$~M$_\odot$ BH turning around an O9.7~Iab companion of $40\pm
10$~M$_\odot$ \citep{Ziolkowski2005} in a circular orbit of 5.6 days
\citep{Brocksopp99a}. High
resolution radio imaging has unveiled the presence of a highly
colimated, relativistic \citep{Stirling2001}, radiatively inefficient
\citep{Gallo2005} jet. The X-ray source displays soft and hard states
and relatively frequent failed transitions between them.  There are
strong evidences of a high energy non-thermal component extending up
to soft gamma-rays \citep{Mcconnell2002,Cadolle2006}. The steady
emission of gamma-rays above 100 GeV is strongly constrained by the
observations with MAGIC which has obtained, however, a very strong
evidence of an intense, fast flaring episode at these energies
\citep{Albert07}.

A $\sim$5.6~d period modulation, attributed to the orbital motion of
the compact object around the companion, has been observed at various
wavelengths by numerous authors
\citep[e.g.,][]{Pooley99,Brocksopp99a,Brocksopp99b,LaSala98,Lachowicz06}.
On the other hand, a super-orbital $\sim$290~d period was claimed by
\citet{Priedhorsky83} on the soft X-ray data recorded by \vela\
(1969-1979) and \ariel\ (1974-1980). Later, a $\sim$150~d periodic
variability has been reported by various authors 
\citep{Pooley99,Brocksopp99b,Ozdemir01,Benlloch01,Benlloch04,Lachowicz06}
using different data samples ranging between April 1991 and November
2003. It must be noted, however, that other significant modulations
with periods $\sim$200~d and $\sim$420~d have been also found
\citep[e.g.][]{Benlloch01,Benlloch04,Lachowicz06}.

In this letter, we search the latest \cyg\ X-ray data for periodic
modulations. We also perform a critical revision of the previous
results obtained from archival X-ray data. Finally, we put our results
in the context of a multiwavelength description of the source.

\section{Data samples} 
\label{sec:data}

\begin{deluxetable*}{lcccccccc}
\tablewidth{0pt}
\tabletypesize{\footnotesize}
\tablecaption{Cygnus X-1 analyzed data\label{tab:data}}
\tablehead{
\colhead{Instrument/} & \colhead{Energy} & \colhead{Operation} & \colhead{Start} & \colhead{End}   & \colhead{Time}     & \colhead{Number of}    & \colhead{\numin}     & \colhead{\numax} \\
\colhead{subsample}   & \colhead{[keV]}  & \colhead{time}      & \colhead{[MJD]} & \colhead{[MJD]} & \colhead{span [d]}  & \colhead{points (m)} & \colhead{[d$^{-1}$]} & \colhead{[d$^{-1}$]} 
}
\startdata
\vela  & 3-12  & May 1969--Jun 1979 & 40368 & 44042 & 3675 & 1097 & 2.7$\times 10^{-4}$ & 0.15 \\
\ariel & 3-6   & Feb 1976--Feb 1980 & 42830 & 44292 & 1464 & 740  & 6.8$\times 10^{-4}$ & 0.25 \\ 
\asm\,1& 2-10  & Sep 1996--Sep 1999 & 50328 & 51444 & 1117 & 913  & 9.0$\times 10^{-4}$ & 0.41 \\
\asm\,2& 2-10  & Jun 2005--May 2008 & 53529 & 54592 & 1064 & 909  & 9.4$\times 10^{-4}$ & 0.43 \\
\swift &15-150 & Jun 2005--May 2008 & 53529 & 54598 & 1070 & 829 &  9.3$\times 10^{-4}$ & 0.39 
\enddata
\end{deluxetable*}

The data samples analyzed in this work are summarized in
Table~\ref{tab:data}.  They are available through the High Energy
Astrophysics Archive Research Center (HEASARC). We use data from four
different instruments, namely: \vela, \ariel\, \asm\ and \swift. All
data are averaged into one-day bins, except when explicitely
stated. No periodic behavior is found in the X-ray data during the
soft state \citep{Wen99,Lachowicz06} and therefore we analyze data
corresponding to the hard state data only. The interval MJD
42338--42829 is dominated by soft flare events \citep{Liang84} and
hence excluded from \vela\ and \ariel\ analyses. The soft state
periods during the operation of \asm\ are identified as those for
which the ratio of count rates in the bands C (5.0-12.1 keV) and A
(1.3--3.0 keV) is lower than 1.2 and the total count rate exceeds the
mean value by more than 4 standard deviations. The mean and standard
deviations are computed from the interval MJD 50660--50990
\citep{Lachowicz06}.  This excludes from the analysis the following
periods (MJD): 50087--50327, 50645--50652, 51002--51026, 51369--51397,
51445--51625, 51776--51952, 52093--52584, 52762--52878, 52982--53092,
53198--53528, 53780--53872. After this, two long intervals dominated
by hard state (samples A and B in Table~\ref{tab:data}) are defined
and studied separately. Based on the results for \asm, the intervals
53414--53528 and 53780--53872 are also removed from the analysis of
\swift\ LC.

\section{Analysis} 
\label{sec:analysis}

We search the different data samples for periodic signals using the
Lomb-Scargle (L-S) test of uniformity \citep{Lomb76,Scargle82}. The
\emph{chance probability} is the probability of obtaining a certain
L-S test value ($z_0$) or larger out of a purely Gaussian noise
sample, and is given by ${\cal P}_\textrm{pre}(z>z_0) =
e^{-z_0}$. When several frequencies are inspected, the {\it
post-trial} probability, i.e. the probability to get a L-S test value
$z_0$ or higher for \emph{at least} one of the scanned frequencies, is
given then by ${\cal P}_\textrm{post}(z>z_0) = 1 - [1-{\cal
P}_\textrm{pre}(z>z_0)]^n$ where $n$ is the number of
\emph{independent} scanned frequencies.

Given $m$ data points, there is a discrete finite set of $m/2$
independent frequencies. For the case of evenly spaced data there is a
natural set of frequencies: $\nu_k = \frac{k}{T}$ ($k=1,\dots,m/2$)
where $T$ is the time spanned by the data set. The values of the
Fourier transform powers for the natural frequencies are independent
of one another. The data set does not contain enough information to
search for periodicities below $\nu_1 = 1/T$ or above $\nu_{m/2} =
m/2T$. The time span, number of data points and the maximum and
minimum accessible frequencies for the different studied data samples
are shown in Table~\ref{tab:data}.

For each investigated data sample we produce the \emph{periodogram},
where $-\log_{10}({\cal P}_\textrm{post})$ is represented as a
function of the frequency. A prominent periodic component in the data
is visible as a peak in the periodogram at the relevant frequency
(\nupeak). We consider as significant those peaks for which the
post-trial probability is lower than $10^{-6.5}$, equivalent to a
deviation of 5$\sigma$ from the Gaussian noise case. We scan all
natural frequencies, with an oversampling factor of 5. This means that
we scan $5m/2$ evenly spaced frequencies, from
\numin\ to \numax. The oversampling does not increase the number of
trials in the post-trial chance probability, since the number of
independent frequencies remains constant, but increases the precision
of \nupeak. To estimate the error ($\Delta$\nupeak), we use the
standard deviation of \nupeak\ over 100 random data samples obtained
by bootstrap \citep{Davison06} of the original LC. Then, we
fold the LC into a
\emph{phaseogram} using the period $1/$\nupeak. The phaseogram is
produced using 50 bins, to ensure a smooth description of the
waveform. The time of the phase 0 ($T_0$) is determined from a fit to
the LC using a Cosine function, where the value of the
frequency is fixed to \nupeak. In this way, $T_0$ corresponds to the
maximum of the fitting Cosine function (although not necessarily to
the maximum of the waveform). The modulation amplitude (\mod) is
defined as the ratio between the amplitude and the mean value of the
Cosine function obtained from the fit. Finally, we remove the periodic
component of frequency \nupeak\ from the LC (a process called
\emph{prewhitening}). This is done by subtracting the deviations of
the phaseogram from its mean value throughout the LC.

We subsequently search for the next most prominent peak in the
prewhitened LC, and follow the whole process described above
in an iterative fashion. This process is stopped when the obtained
\nupeak\ has a post-trial chance probability larger then $10^{-6.5}$.

\section{Results}   

\begin{figure}
\epsscale{1}
\plotone{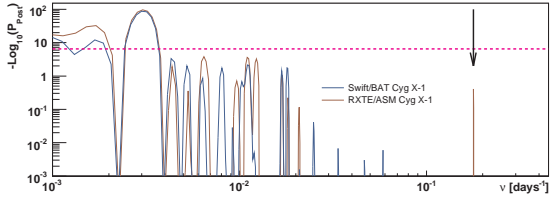}
\figcaption[f1.eps]{Periodograms for \swift\ and
\asm\,2 \cyg\ samples, showing the post-trial chance probability as a
function of the scanned frequency. The horizontal line marks the line
corresponding to a post-trial probability of $10^{-6.5}$. The arrow
marks the orbital frequency.
\label{fig:asm_swift_per}}
\end{figure}

\begin{figure}
\epsscale{1}
\plotone{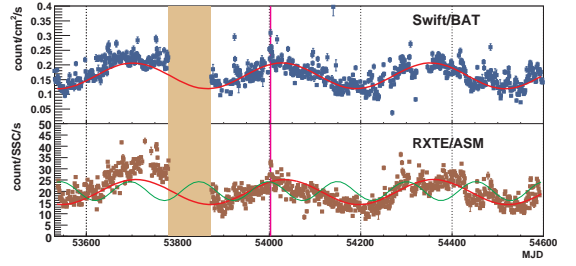}
\figcaption[f2.eps]{\swift\  and \asm\,2 LCs. The
shaded area shows an interval of soft state, identified by the
criteria exposed in Section~\ref{sec:data}, and not considered in the
analysis. The thick, red curves are the fits by Cosine functions to
each subsample (see Section~\ref{sec:analysis}). The thin, green curve
represents the $\sim 150$~d super-orbital modulation using the
ephemeris reported by
\citet{Lachowicz06}. The vertical, purple line marks the time of the
TeV signal reported by
\citet{Albert07}
\label{fig:asm_swift_lc}}
\end{figure}

\begin{figure}
\epsscale{1}
\plotone{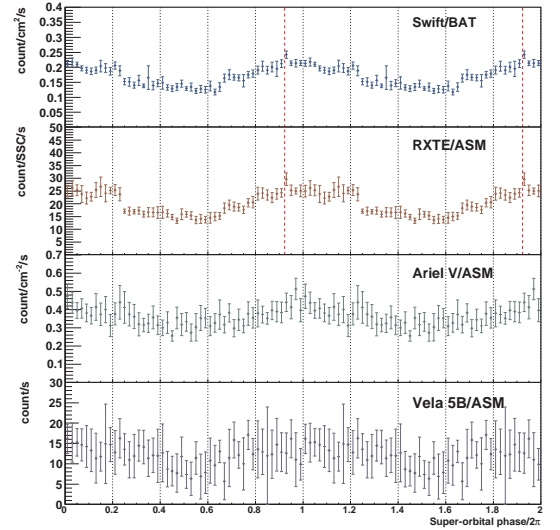}
\figcaption[f3.eps]{From top to bottom:
\swift, \asm\,2, \ariel\ and \vela\ phaseograms folded using 
period P and time 0 values (P [d],$T_0 \textrm{[MJD]}$)= (326,
54027), (326, 54027), (276, 42866) and (288, 40187), respectively.
The values of $T_0$ are obtained from the fit of a Cosine
function. Data points and error bars correspond, respectively, to the
mean count rate and variance measured within each phase bin. The
vertical, red, dashed line corresponds to the phase of the TeV signal
reported by \citet{Albert07}. \label{fig:phaseogram}}
\end{figure}

We first search for periodic signals in \swift\ and \asm\,2 data
samples, which correspond to the same epoch, and which are analyzed in
this work for the first time. The results are shown in
Table~\ref{tab:results}. The L-S periodograms for both LCs are shown
in Figure~\ref{fig:asm_swift_per}. A very strong, dominant periodic
signal with period 326$\pm$2~d is found in both data samples. The LCs
are shown in Figure~\ref{fig:asm_swift_lc}. The modulation is clearly
seen by eye, which is reflected by the extremely low values of
\postrial. The previously reported $\sim$150~d modulation is not found
in these data. The pre-trial chance probabilities for such a
modulation are $10^{-1.3}$ and $10^{-0.1}$ for \asm\ and
\swift\ data, respectively. Figure~\ref{fig:asm_swift_per} shows the
$\sim 150$~d modulation as reported by \citet{Lachowicz06} overlaid with
\asm\,2 LC, confirming that such a modulation does not describe well the
data. The phaseograms corresponding to the 326~d period are shown in
Figure~\ref{fig:phaseogram}. We see that hard and soft X-ray LCs are
strongly correlated, with Pearson's correlation factor $r=0.97$. A
second, also strong, component with period $\sim$1000~d is present in
both data samples. This corresponds to a long-term modulation of the
X-ray flux with respect to the 326~d oscillation, but cannot be
established as periodic since the period is similar to the total time
spanned by the observations. An alternative explanation to the
$\sim$1000~d period will be given below. Finally, a third component is
seen in \asm\,2 data sample at period $\sim$5.6~d, compatible with the
orbital modulation, visible in the periodogram
(Figure~\ref{fig:asm_swift_per}) even before prewhitening. This
modulation is not seen in \swift\ data, for which ${\cal
P}_\textrm{pre} \simeq 10^{-2}$. On a similar energy band,
\citet{Paciesas97} claimed a modulation compatible with the orbital
period in \batse\ LC between April 1991 and September 1996, but the
method used lacks of a mathematical
justification. \citet{Brocksopp99b} did not find any evidence for the
orbital modulation in \batse\ data between May 1996 and September
1998. Finally, \citet{Lachowicz06}, using the whole BATSE light curve,
reported a deviation of $\sim 1\sigma$ from the Gaussian noise case,
insufficient to establish the presence of the orbital modulation in
the data. 

We have searched \vela\ and \ariel\ LCs for periodic
modulations and found peaks at P=(276$\pm$3)~d and P=(288$\pm$3)~d,
respectively (see Table~\ref{tab:results}), in agreement with the
results obtained by \citet{Priedhorsky83} and
\citet{Lachowicz06}. By comparison with the P=(326$\pm$3)~d present in
\swift and \asm\,2 LC, this shows that the period of the super-orbital
modulation is variable. The corresponding phaseograms are shown if
Figure~\ref{fig:phaseogram} (two lowermost panels). Even if the
relative dispersion of the points is larger due to the lower
sensitivity of \ariel\ and \vela, the waveform is still visible. They
follow a very similar shape as those found for \swift\ and \asm\,2,
albeit for a different period, as one expects if the underlying
physical process is the same. The correlation factor for \swift\ and
\ariel\ (\vela) phaseograms is $r=0.73$ ($r=0.57$). However, the
modulation amplitudes are significantly lower than for the case of
\asm. This could have a physical explanation, but it could also happen
if the periodic modulation was not present in part of the LCs, which
can be certainly not excluded. On the other hand, we do not find
evidence for the orbital period in
\vela\ or \ariel\ LCs\footnote{We note that shorter integration times
have been used for this search in the
\vela\ LC since, with 1-day bins, the minimum
accessible period is P=6.7~d, as shown in Table~\ref{tab:data}}. It
is worth noting that, given \vela\ and \ariel\ sensitivities, we do
not expect to detect an orbital modulation with an amplitude of
$\sim$5$\%$ as the one we see in \asm\,2 data. The mean relative
variance of the data points in the phaseogram (which is a good
estimate of the measurement error) are $46\%$ and $13\%$ for
\vela\ and \ariel\, respectively. Both values are well above the $5\%$
modulation which is hence difficult to detect. We have crosschecked
this by analyzing Monte Carlo (MC) simulated LCs for \vela\
and \ariel. We use the same sampling as the measured LCs and
simulate a $5\%$ amplitude modulation convolved with $46\%$ and $13\%$
point-to-point random fluctuations, respectively. The analysis of
these LCs yields no significant peak.

\begin{deluxetable}{lr@{$\pm$}lrrl}
\tablewidth{0pt}
\tablecaption{Results for periodic signal search\label{tab:results}}
\tablehead{
\colhead{Sample} & \multicolumn{2}{c}{P$_\textrm{peak}$} & \colhead{$T_0$} &  \colhead{\mod} & ${\cal P}_\textrm{post}$ \\ 
                 & \multicolumn{2}{c}{[d]}               & \colhead{[MJD]} &  \colhead{[$\%$]} &
}
\startdata
\swift   & 326   &  2    & 54027.3 & 25  &  $10^{-87}$  \\
         & 1030  & 50    & 53707.2 &  6  &  $10^{-10}$  \\
\asm\,2  & 326   & 2     & 54032.2 & 29  &  $10^{-96}$  \\
         & 990   & 30    & 53670.7 & 13  &  $10^{-41}$  \\
         & 5.600 & 0.002 & 53670.3 &  5  &  $10^{-11}$  \\
\asm\,1a & 248   & 9     & 50377.2 & 29  &  $10^{-30}$  \\
\asm\,1b & 123   & 3     & 50761.1 & 11  &  $10^{-10}$  \\
\asm\,1c & 168   & 4     & 51177.6 & 18  &  $10^{-24}$  \\
\asm\,1  & 5.602 & 0.002 & 51117.1 & 4.1 &  $10^{-7.1}$ \\
\ariel   & 276   & 3     & 42865.8 & 14  &  $10^{-16}$  \\
\vela    & 288   & 3     & 40187.4 & 21  &  $10^{-6.8}$
\enddata
\end{deluxetable}

\begin{figure}
\epsscale{1}
\plotone{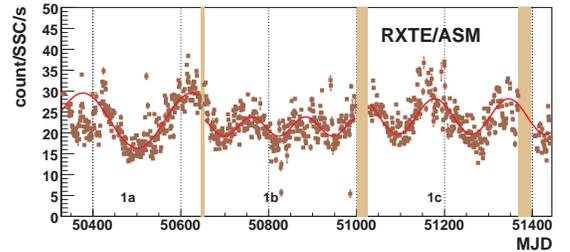}
\figcaption[f4.eps]{\asm\,1 light curve. The vertical
shaded lines show the intervals of soft or failed transition states,
identified by the criteria exposed in Section~\ref{sec:data}, which
delimit the three subsamples (1a, 1b and 1c) used for further
analysis. The red curves are the fits by Cosine functions to each
subsample (see Section~\ref{sec:analysis}).
\label{fig:asmA_lc}}
\end{figure}

Finally, we have searched \asm\,1 data sample for periodic
modulations, and found 5 significant peaks at P = (148$\pm$1),
(188$\pm$2), (310$\pm$11), (475$\pm$11) and (5.598$\pm$0.003)~d, all
with chance probabilities lower than $10^{-10}$. The latter
corresponds to the orbital modulation, whereas the other four seem to
denote a complex power spectrum. We stress that some of these peaks
have been found in previous studies of the \asm\ LC
\citep{Benlloch01,Benlloch04,Lachowicz06}. The understanding of the
super-orbital modulation can be greatly simplified if we consider that
the period can change along the observation time in a discrete way. We
have analyzed separately the data between each two consecutive soft or
failed transition states, i.e.\ three samples, namely
1a=[50328--50644], 1b=[50653--51001] and 1c=[51027--51368] (see
Figure~\ref{fig:asmA_lc}). We obtain a single significant peak in each
of them (see Table~\ref{tab:results}). We prewhiten the leading
frequency in each of the subsamples and merge them to look for
sub-leading frequencies. The only remaining significant peak
corresponds to the orbital modulation (see
Table~\ref{tab:results}). Figure~\ref{fig:asmA_lc} shows \asm\,1 LC,
overlaid with the result of the fits to independent Cosine functions
to the three defined subsamples. The general agreement with the data
is remarkably good. We have also generated three MC samples
corresponding to the samplings of 1a, 1b and 1c subsamples, and pure
sinusoidal modulations with the periods found for each of them,
convolved with the measured point-to-point fluctuations. Then we have
merged the three samples together and analyzed the resulting LC. We
obtain significant peaks at P = (147$\pm$1), (190$\pm$2) and
(523$\pm$12)~d, in surprisingly good agreement with the results of
analyzing the real data. We note that this effect could be also
responsible of the $\sim$1000~d periodicity of the \swift\ and \asm\,2
data, since they also contain a soft state episode that might have
changed the period of the super-orbital modulation prior to MJD=53780.

\section{Discussion}   

We find that \cyg\ displays a super-orbital modulation, with a period
that changes, probably in a discrete way and in coincidence with soft
or failed state-transition phases, over time scales ranging from a few
hundred days to several years. According to our findings, the very
much cited $\sim$150~d period is most probably an artifact of applying
a (sometimes biased) Fourier-transform based analysis to a data sample
where more than one consecutive period modulations are present. Since
2005, \cyg\ shows a very powerful and stable super-orbital modulation
with a period of $326\pm2$~d.

The super-orbital modulation is usually attributed to the precession of
the accretion disk~\citep{Priedhorsky83} and relativistic
jet~\citep{Romero02}, as a result of the tidal forces exerted by the
companion star on a tilted disk~\citep{Katz73}. A mechanism for
keeping the disk tilted can be provided by radiation pressure
warping~\citep{Petterson77,Pringle96,Wijers99,Ogilvie01}. In the
case of tidally forced precession, the expected period
$P_\textrm{prec}$ depends on the outer radius $R_o$ and inclination of
the disk $\delta$ as $P_\textrm{prec} \propto R_o^{-3/2}
\cos^{-1}\delta$ \citep{Larwood98}. Then, the longer the period the
larger the precession angle, and hence also a larger modulation
amplitude is expected. This is in agreement with our results for
\asm, where we have found four different super-orbital periods, which
follow this tendency (see Table~\ref{tab:results} and
Figure~\ref{fig:asmA_lc}).

A different issue is why the precession of the disk produces a
modulation in the X-ray flux. Some authors
\citep[e.g.,][]{Lachowicz06,Ibragimov07} have considered the
possibility that the precession movement changes the optical thickness
along the line of sight. They reject this possibility since it seems
unlikely due to the fine tuning required to produce the observed
modulation amplitude, which in addition should depend on the energy,
which is not confirmed by our observations. The multiwavelenth data
seem to support a scenario where the emission itself is
anisotropic. The precession modulation is detected at similar times
with identical periods in radio, soft and hard X-rays during the hard
state. A unified picture, where the anisotropy is provided by the jet,
has been proposed by \citet{Brocksopp99b}. The soft X-ray emission is
produced in the disk via bremsstrahlung of thermal electrons, and are
then up-scattered to higher energies via Compton scattering in the hot
corona or at the base of the relativistic jet (which precesses with
the disk). The acceleration of electrons along magnetic field lines in
the jet produces the radio emission by synchrotron emission. During
the soft state, the jet and corona dissappear and no modulation is
observed. According to our findings, once the source goes back to the
hard state, the reconstructed disk and jet have different kinematical
properties, and the modulation period changes. It seems that failed
transitions produce a similar effect.

MAGIC detected a fast and intense episode of emission of gamma-rays
above 100~GeV \citep{Albert07} during MJD=54003, albeit at the limit
of the telescope's sensitivity. This happened in coincidence with the
soft and hard X-ray maxima (Figures~\ref{fig:asm_swift_lc} and
\ref{fig:phaseogram}) and an unusually bright outburst detected with
{\it INTEGRAL} \citep{Malzac08}. It is interesting to note that,
according to the ephemeris shown in Table~\ref{tab:results}, the next
passage for the precession maximum will happen at MJD=54655$\pm$2,
i.e. between 6 and 10 of July 2008. The observational conditions of
\cyg\ with ground-based Cherenkov telescopes, such as MAGIC and
VERITAS, will be optimal during those days.

\acknowledgments I would like to warmly thank the help, discussions and comments
to the draft from Emma de O\~na-Wilhelmi, Roberta Zanin, Diego Torres,
Daniel Mazin, Juan Cortina and Miguel A. P\'erez-Torres.



\end{document}